\documentclass[twocolumn]{article}
\usepackage{xcolor} \usepackage[super]{natbib}
\setcitestyle{comma}

\usepackage{amsmath}
\usepackage{authblk}
\usepackage[top=1.5cm,left=1.5cm,right=1.5cm,bottom=2cm]{geometry}

\usepackage{graphicx}
\usepackage{txfonts}
\usepackage[separate-uncertainty]{siunitx}
\DeclareSIUnit{\mearth}{M_\oplus}
\DeclareSIUnit{\mjup}{M_{Jup}}
\DeclareSIUnit{\rearth}{R_\oplus}
\DeclareSIUnit{\rjup}{R_\jupiter}
\DeclareSIUnit{\rsol}{R_{\odot}}
\DeclareSIUnit{\msol}{M_{\odot}}
\DeclareSIUnit{\au}{au}
\DeclareSIUnit{\year}{yr}
\DeclareSIUnit\erg{erg}
\usepackage{graphicx}
\usepackage{ulem}

\usepackage[]{hyperref}               
\hypersetup{pdfauthor=R. Burn et al.}
\hypersetup{breaklinks=true,colorlinks=true,urlcolor=blue, linkcolor=blue,  citecolor=blue,anchorcolor=blue,bookmarksopen=true}

\graphicspath{{./fig/}}

\definecolor{snsgreen}{rgb}{0.0, 0.620, 0.451}

\title{
A radius valley between migrated steam worlds and evaporated rocky cores}
\author[1]{Remo Burn}
\author[2]{Christoph Mordasini}
\author[3]{Lokesh Mishra}
\author[2]{Jonas Haldemann}
\author[4]{Julia Venturini}
\author[5]{Alexandre Emsenhuber}
\author[1]{Thomas Henning}
\affil[1]{Max-Planck-Institut f\"ur Astronomie, K\"onigstuhl 17, Heidelberg, 69117, Germany}
\affil[2]{Physikalisches Institut, Universit\"at Bern, Gesellschaftsstrasse 6, Bern, 3012, Switzerland}
\affil[3]{IBM Research, R\"uschlikon 8803, Switzerland}
\affil[4]{Observatoire de Gen\`eve, Chemin Pegasi 51, Versoix, 1290, Switzerland}
\affil[5]{Universit\"ats-Sternwarte M\"unchen, Ludwig-Maximilians-Universit\"at M\"unchen, 
	Scheinerstra\ss{}e 1, M\"unchen, 81679, Germany}

\begin{document}

\twocolumn[
\begin{@twocolumnfalse}
	\maketitle
	\begin{abstract}
	    \textbf{The radius valley (or gap) in the observed distribution of exoplanet radii, which separates smaller super-Earths from larger sub-Neptunes \citep{Fulton2017,Fulton2018,Hsu2019}, is a key feature that theoretical models must explain. Conventionally, it is interpreted as the result of the loss of primordial H/He envelopes atop rocky cores \citep{Owen2013,Lopez2013,Jin2014,Owen2019a}. 
		However, planet formation models \citep{Emsenhuber2020a,Emsenhuber2020b,Venturini2020} predict that water-rich planets migrate from regions outside the snowline toward the star. Assuming water to be in the form of solid ice in their interior, many of these planets would be located in the radius gap\citep{Jin2018}, in disagreement with observations. Here we use an advanced coupled formation and evolution model that describes the planets' origin and evolution starting from moon-sized, planetary seed embryos in the protoplanetary disk to mature Gyr-old planetary systems. Employing new equations of state and interior structure models to treat water as vapor mixed with H/He, we naturally reproduce the valley at the observed location. The model results indicate that the valley separates less massive, in-situ, rocky super-Earths from more massive, ex-situ, water-rich sub-Neptunes. Furthermore, the occurrence drop at larger radii, the so-called radius cliff\citep{Kite2019}, is also matched by planets with water-dominated envelopes. Owing to our statistical approach, we can assess that the synthetic distribution of radii quantitatively agrees with observations for the close-in population of planets; but only if atmospheric photoevaporation is also acting, populating the super-Earth peak with evaporated rocky cores. Therefore, we provide a hybrid theoretical explanation of the radius gap and cliff caused by both formation (orbital migration) as well as evolution (atmospheric escape).
		}
	\end{abstract}
  \end{@twocolumnfalse}
]

The distribution of planetary radii observed mainly by the Kepler spacecraft shows a prominent feature called the ``radius valley'' \citep{Fulton2017,Fulton2018,Hsu2019,Ho2023}. It separates smaller ``super-Earths'' with radii $R \lesssim\SI{1.7}{\rearth}$ from larger ``sub-Neptunes'' and its origin has multiple interpretations. One is a planetary envelope mass-loss process, i.e. atmospheric escape during planet evolution. The driving process could be photoevaporation because of stellar X-ray and Ultraviolet irradiation\citep{Owen2013,Lopez2013,Jin2014,Owen2019a} or core-powered mass loss \citep{Ginzburg2016,Owen2016a,Rogers2021} of a hydrogen-helium (H/He) envelope on top of a core consisting of silicates and iron. In the following, we call this mixture of solid materials ``rocky''. With this limited compositional inventory, the observed sharp drop-off of the sub-Neptune occurrence at radii greater than \SI{\sim3}{\rearth} -- the ``radius cliff'' -- likely requires an additional mechanism, such as H$_2$ sequestration in the magma ocean \citep{Kite2019}.

Another possible explanation for the radius distribution is that the sub-Neptunes are water-rich with water mass fractions of several tens of percent \citep{Zeng2019,Mousis2020,Aguichine2021}. Such large water contents are a consistent prediction of planet formation models which include the effect of planet-disk interactions leading to migration of ice-rich planets from outside the water condensation line towards the star \citep{Ward1997,Mordasini2009b,Ida2013,Bitsch2019,Brugger2020,Venturini2020}. In this scenario, the commonly assumed H/He dominated envelope of the sub-Neptunes is not required anymore. This hypothesis has recently gained supporting observational evidence based on planets around M stars \citep{Luque2022}, whose bulk density distribution can be well reproduced with silicates and iron for super-Earths and about equal fractions of rock and water for sub-Neptunes. The tentative evidence for this scenario lies in the small scatter of sub-Neptune densities \citep{Luque2022}, which is not expected for rocky cores with a H/He envelope.

For more massive stars, the smaller radial velocity amplitude often hinders a precise determination of the planetary mass. In that case, a clear picture in density does not emerge with present-day data and the different theoretical models can not easily be falsified \citep{Rogers2021a} unless the planetary masses are constrained using theoretical arguments, such as the output of a planet formation model \citep{Jin2014,Jin2018,Venturini2020,Izidoro2022}. 

In the case of water-dominated outer layers of the planet, the phase of water determines the precise radius. One key assumption, which is well motivated at the high temperatures where typical sub-Neptunes are observed, is that water is not condensed out into solid ice and instead forms a massive, hot, to a large degree supercritical, vaporized hydrosphere, here called ``steam envelope'' \citep{Mousis2020}. This significantly increases the radius relative to the condensed case \citep{Turbet2020}. Due to model limitations, water was not consistently included in this lower-density phase in the first works coupling a planetary mass distribution from planet formation models to photoevaporation models \citep{Jin2014,Jin2018} and/or comparing the observed valley locus with the theoretically predicted one \citep{Owen2017,Jin2018}. Therefore, these authors excluded that the super-Earths contain significant amounts of water and similarly favored water-poor sub-Neptune compositions because for solid ice, such planets fill the radius gap.
Later, a combined formation and evolution model with the correct water phases to show that the the radius valley can emerge as a separator between dry and wet planets (i.e, formed within vs. beyond the ice-line) was presented \citep{Venturini2020}. In that work, the main process driving such a dichotomy are different efficiencies for pebble accretion depending on the pebble composition, which produces smaller rocky and larger icy cores. Recently, another study \citep{Izidoro2022} came to the same conclusion with a similar model but also modeling \textit{N}-body interactions between planets while keeping the water in condensed form.
In ref. \citep{Venturini2020}, the authors found that the location of the valley is better reproduced if any water is mixed with H/He as steam and both undergo atmospheric escape. A detailed quantitative comparison to observations and the simulation of \textit{N}-body interactions was outside the scope of their work motivating further investigation.

Here, we used a planetesimal-based coupled formation and evolution model to synthesize a population of planets. For the nominal case, we assumed that any accreted water mixes with H/He and can form a steam envelope. Furthermore, we model photoevaporative mass loss of both water and H/He leading to a more complete picture of all processes at play. The results are obtained in a statistically robust way by modeling a full population of 1000 systems with initial conditions informed from observations of disks \citep{Tychoniec2018,Andrews2018,Richert2018}. This enables us to extract the key mechanism shaping the radius valley and statistically quantifying their ability to reproduce observations.

We use the results of our planet formation modeling \citep{Emsenhuber2020a,Emsenhuber2020b} created for statistical studies. The planetary systems form around \SI{1}{\msol} stars, each of which starts with different disk properties and initially 100, randomly distributed, \SI{0.01}{\mearth} planetary seed particles per disk. During the formation stage we model the evolution of the gas disk, planetary growth by accretion of planetesimals and gas, gas-driven planetary migration, and dynamical interactions between the planets by means of an \textit{N}-body algorithm (scattering, giant impacts, and -- combined with orbital migration -- capture into resonances). The formation model is outlined in the Methods section.

In the subsequent evolution stage, the planets are evolved individually, taking into account in the calculation of their interior structure the effects of cooling and contraction, atmospheric photoevaporation, bloating, and stellar tides. This leads to a population of planets at an age of \SI{5}{\giga\year}.

For the results shown here, we have extended the formation phase to \SI{100}{Myr} compared to 20 Myr in the original work \citep{Emsenhuber2020b} before evolving the individual planets. For the nominal model, we mix the accreted water with any present H/He \citep{Vazan2022} and, for water, use a new equation of state \citep{Haldemann2020}, which covers all possible phases. Under this assumption, water is also present in the upper layers of the gaseous envelope and can even be the only volatile constituent. Therefore, we include the photoevaporative mass loss of water \citep{Johnstone2020} and add it mass weighted to the photoevaporative loss of H/He \citep{Kubyshkina2021}. The evolution processes and their implementation are also detailed in the Methods section.

As a final step, in order to compare synthetic results to transit observations, we apply the detection biases from the Kepler mission using KOBE \citep[][]{Mishra2021}. The procedure is described in the Methods and allows for a statistical comparison to the California Kepler Survey \citep{Petigura2017} supplemented with Gaia data \citep{Fulton2018}.

\section*{Results}
\begin{figure}
	\centering
	\includegraphics[width=\linewidth,trim=0 7 0 7, clip]{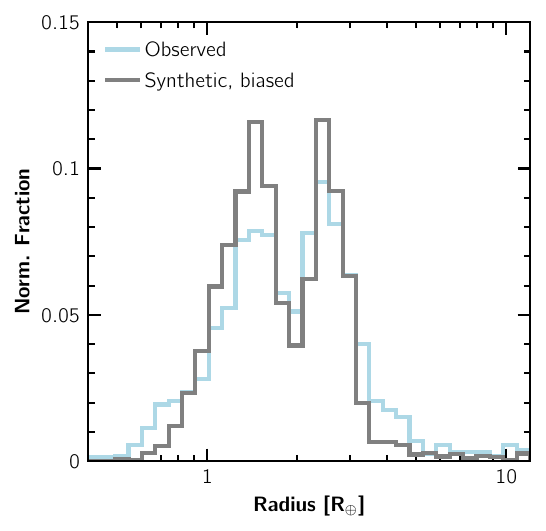}
	\caption{Radius histograms of observed and synthetic planets. The light blue line shows the observed distribution without correction for the observational bias \citep{Fulton2018} and the gray line the synthetic one using the updated, nominal evolution model with the bias of the Kepler survey applied. We restrict the sample to planets with orbital periods lower than \SI{100}{days}. Bin counts are normalized by the total number of planets in the different samples.}
	\label{fig:r_hist_fiducial}
\end{figure}
\begin{figure*}
	\centering
	\includegraphics[height=.37\linewidth,trim=10 7 7 0]{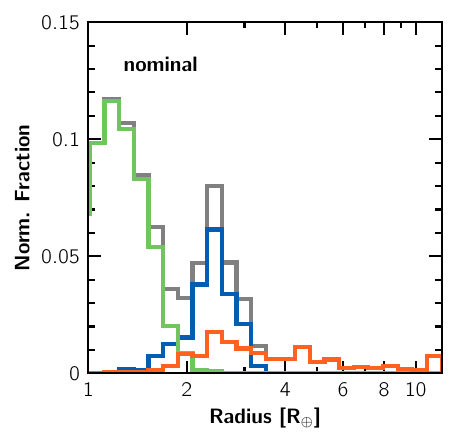}
	\includegraphics[height=.37\linewidth,trim=40 7 7 0 , clip]{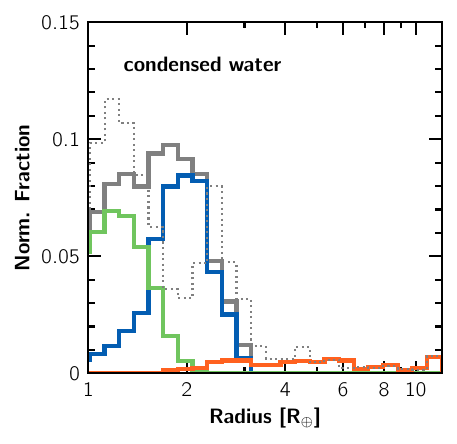}
	\includegraphics[height=.37\linewidth,trim=40 7 7 0 , clip]{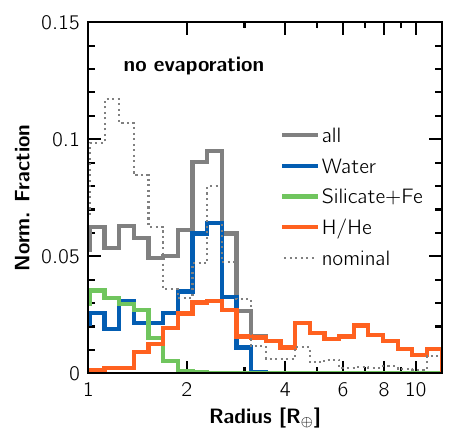}
	\caption{Radius histograms of synthetic planets with orbital periods shorter than \SI{100}{days} for different model assumptions without any bias applied. For each panel, the full distribution is shown in gray and colored histograms are showing different compositional subsets (pure rocky in green, water-rich without any H/He in blue, with some H/He in red). The left panel shows the nominal model, in the middle panel water was assumed to be condensed out in a solid ice layer resistant to evaporation below H/He, and the right panel uses the nominal treatment of water but excludes evaporation. The dotted histogram shows the full distribution from the nominal simulations. As in Fig. \ref{fig:r_hist_fiducial}, bin counts are normalized by the total number of planets in the different samples.
	}
	\label{fig:r_hist_compo}
\end{figure*}
\paragraph{Radius distribution}
\label{sec:results_main}
We contrast our theoretical radius distribution with applied observational bias to observations\citep{Fulton2018} in Fig. \ref{fig:r_hist_fiducial}. We find an excellent match for the locus of the sub-Neptune peak, the radius valley, and the super-Earth peak. The synthetic radius cliff is only marginally steeper while the relative number of super-Earths in the model is higher than observed. Statistical quantification of the differences (Extended Data Fig. \ref{fig:statistics}) reveals that the radius distribution of super-Earths within \SI{30}{d} is well matched but at larger distances, we synthesize and theoretically ``detect'' more rocky planets than observed. The depth of the radius valley is deeper in the synthetic population, which is in this comparison not affected by a smoothing effect of measurement uncertainties as for the observed distribution\citep{Ho2023}. We note that the match to the observed distribution of planetary radii was obtained naturally from consistently including more realistic physics, especially the different phases of water, compared to previous works \citep{Jin2018}.  The result is obtained from following the self-consistently formation and evolution of initially 100 lunar-mass planets per disk. We did not adjust any parameters neither in the underlying formation phase nor in the evolutionary model.

In Fig. \ref{fig:r_hist_compo}, we show radius distributions of different modeling runs without an observational bias applied, splitting the population according to bulk composition. From the left panel in Fig. \ref{fig:r_hist_compo}, it can be inferred that the valley separates smaller, dry planets from lower-density, wet planets with a vapor envelope. There is only a small overlap of the two distributions. The planets whose envelopes contain some H/He make up only 13\% to 16\% (depending weakly on age) of the sub-Neptunes. Furthermore, most of the planetary envelopes of sub-Neptunes with some H/He, contain less than ten percent of H/He by mass.

The radius cliff located at \SI{\sim 3}{\rearth} is in our simulations a second compositional transition from planets containing $\gtrsim$\SI{90}{\percent} water to those which contain several tens of percent of H/He in mass. It is accompanied with a corresponding drop in the occurrence of planets at \SI{\sim 20}{\mearth}. Thus it is a feature which emerges due to more gas accretion for the more more massive planets.

We further varied model assumptions to understand how robust the results are. We found the results to be insensitive to any bloating mechanism (Supplementary Information). In contrast, under the assumption that water is buried below the H/He envelope and is forced to remain in condensed form we do not reproduce the radius valley (middle panel of Fig. \ref{fig:r_hist_compo}). Instead, water-rich cores populate the radius range where the observed valley is located. This echoes previous works \citep{Owen2017,Jin2018} excluding water-rich cores and shows that the cause of recovering a radius valley with water-rich planets can be attributed to the (correct) phase of water and its distribution within the envelope \citep{Mousis2020}.

The rightmost panel of Fig. \ref{fig:r_hist_compo} shows the case of excluding photoevaporation while keeping the water mixed into the H/He envelope. There, we get a distribution which is neither in agreement with the presence of a radius valley. Instead, we obtain many (low-mass) large planets with a rocky core and a H/He envelope and also a distribution of water-rich planets which smoothly extend to low radii. In reality, both of these kinds of planets would be unable to retain their H/He or (to a smaller extent) water envelopes. For the same reason, too few rocky planets exist. This highlights the need of atmospheric escape shaping the distribution of planetary radii even for water-rich compositions. It is necessary to populate the super-Earth peak with rocky planets by stripping their H/He envelopes. We conclude that the valley is a hybrid consequence of both formation (migration leading to the sub-Neptune peak) and evolution (evaporation leading to the super-Earth peak).

\paragraph{Dependency on orbital period}
\label{sec:results_slope}

\begin{figure*}
	\centering
	\includegraphics[width=\linewidth,trim=60 5 80 30, clip]{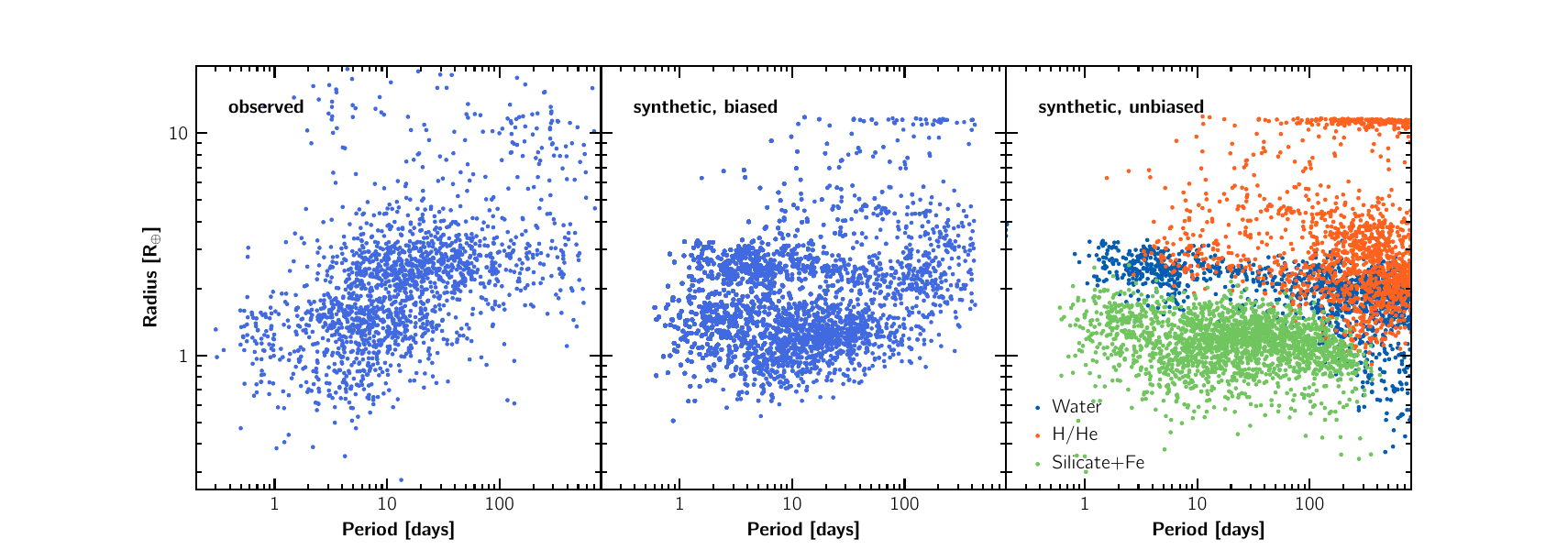}
	\caption{Transit radii as a function of orbital period of observed (left) and the nominal, synthetic population. The data for the observational scatter plot is from the California Kepler Survey \citep{Petigura2018}. The synthetic data is biased (middle panel) using KOBE (see Methods). The rightmost panel shows the unbiased synthetic distribution color coded by their composition (as in Fig. \ref{fig:r_hist_compo}).
	}
	\label{fig:pR_vanilla}
\end{figure*}

In addition to radii, orbital periods of exoplanets can be determined precisely. Fig. \ref{fig:pR_vanilla} shows the period-radius distribution of observed \citep{Fulton2018} (left) and modeled (middle and right) exoplanets. Qualitatively, the observed distribution matches well the synthetic distribution with applied observational bias. We note that we apply the bias of the full Kepler survey without taking into account that not all planets are included in the California Kepler Survey catalog \citep{Fulton2018}. Therefore, a larger number of points is shown in the middle panel. In both the observed and the synthetic distribution, the radius valley can be made out at comparable locations.

The right panel shows the synthetic data without observational bias applied. Additionally, the composition is color-coded. Rocky Earths and super-Earths (green) are found at small radii below the valley. This distribution is truncated at around \SI{300}{d} where the equilibrium temperature is close to \SI{300}{K}.

The planets with water but without H/He (blue) are filling the parameter space at radii greater than the rocky planets above the valley and at larger distances. Planets with H/He (red) populate the even larger radii and are also more common at lower radii further away from the star.

As revealed by the selected formation tracks in Extended Data Fig. \ref{fig:pM_vanilla}, this pattern is shaped by migration and collisions during the formation stage as well as photoevaporation. Rocky planets generally form at short distances inside of the water snowline. They grow first by planetesimal accretion and then (and most importantly) by giant impacts with other rocky protoplanets. Giant impacts are visible as vertical parts in the formation tracks. However, their mass growth is limited due to the limited amount of building blocks inside of the ice-line. In absolute terms, only little orbital migration inside of the ice-line happens.

At larger distances to the star, more solids are available for accretion for the assumed MMSN-like profile of the planetesimal disk. In addition, even more material is available for solid accretion thanks to the condensed icy material outside of the water ice-line. This leads to promoted planetary growth up to a several Earth masses where the timescale of type I migration \citep{Paardekooper2011} reduces and planets migrate efficiently into the inner region. While migrating, the more massive, water- and H/He-rich planets interact and often collide with rocky planets in the inner system, especially after the systems are destabilized after the gas disk dissipates due to a lack of eccentricity damping\citep{Cresswell2008}. Collisions with other bodies -- or at the closest distances the radiation flux of the host star -- can lead to the expansion and Roche lobe overflow of the gaseous envelope and thus remove the H/He content. However, under our assumptions, this is not the case for the heavier water. Only after switching to the evolution stage, we assume that the constituents perfectly mix. These processes therefore give rise to a population of planets with pure water envelopes. More massive, larger planets or planets at larger distances share a similar formation history but are not stripped completely of H/He (red planets and tracks in Fig. \ref{fig:pR_vanilla} and Extended Data Fig. \ref{fig:pM_vanilla}).
 
During the evolution stage, planets cool and are subject to photoevaporation. Thus, they move vertically downwards in Fig. \ref{fig:pR_vanilla}. The most frequent evolutionary pathway is the loss of a (pure) H/He atmosphere of a rocky core. In a smaller fraction of the cases, also a mixed or water-dominated envelope can be lost completely to result in a bare rocky core. This happened for \SI{17}{\percent} of the (eventually) rocky super-Earths in the biased, synthetic population which had at least ten percent of water by mass after the formation stage. This rare outcome occurs for the lowest mass planets which were able to migrate.

Although there is an overlap in mass space between volatile-rich and rocky planets (see Extended Data Fig. \ref{fig:mass_hist}), the overall formation pathway occurs along the following lines: rocky cores are lower mass planets which formed almost in-situ by a giant impact stage while volatile-rich planets are more massive and for that reason migrated substantially to their present-day location. Low-mass volatile-rich in contrast do not migrate close to the host star in significant numbers (which would fill the valley), because type I migration is slower the lower the planet mass is\citep{Ward1986}.

\section*{Discussion}
\paragraph{Mass distribution and mass-radius diagram}

A fundamental property of planets with regards to their formation is their mass. Here, we started at planetary embryos of \SI{0.01}{\mearth} and modeled their growth by solid and gas accretion as well as the long-term evolution of the bodies which results in a distribution of planetary masses shown in Extended Data Fig. \ref{fig:mass_hist}. This approach is very different from mass distributions derived from fitting the observed properties of the close-in low-mass population found by the Kepler satellite via a inference analysis \citep{Gupta2020,Rogers2021a}.
As mentioned above, there is a clear trend in planetary mass from low-mass, rocky planets over more massive, migrated, steamy water worlds to the H/He-rich (sub)-giants.  However, the mass distributions are broad enough to lead to a significant overlap which results in a unimodal total mass distribution.

The characteristic masses are few Earth masses for rocky planets and about ten Earth masses for water worlds, but with broad distributions. These mass scales can be understood analytically by recalling that most of the rocky planets went through a giant impact phase after a destabilization of the systems. The appropriate mass scale is then given by the amount of solid building blocks in an dynamically enhanced feeding zone (Goldreich mass) \citep{Goldreich2004,Weiss2022PPVII,Emsenhuber2023a}. For planets which migrate, thus the water-rich planets, the equality of migration against growth or -- if present -- the saturation of co-rotation torques set the mass scale (Equality or Saturation mass) \citep{Weiss2022PPVII,Emsenhuber2023a}.

While the mass of the rocky planets depends on the solid accretion mechanism, the mass of migrated steam-worlds is less sensitive to it. Independently from our work, the study \cite{Venturini2020} using a global model with pebble accretion instead of planetesimal accretion resulted in a mass distribution with more distinct separation of rocky and water-rich planets in mass. An overall similar mass distribution to ours was retrieved in a Bayesian framework by \citep{Rogers2021a} shown in Fig. \ref{fig:mass_hist} (dashed). A strong difference is however that a very low probability of planets containing water was inferred which we attribute to their work assuming water to be in the icy/condensed phase at all temperatures.

In Extended Data Fig. \ref{fig:mR}, we also show the synthetic mass-radius diagram for the nominal model at 5 Gyr. We find that our model leads to planets covering the area occupied by precisely characterized, observed planets around solar-type stars. 
A tentative over-density of both observed and simulated planets is located close to the lowest-density planets without H/He (uppermost blue points in the diagram). This would be in agreement with recent observational findings \citep{Luque2022} for M stars but is for Solar-type stars without statistical significance. So far, no unbiased, statistical sample of characterized planets with precise mass and radius measurements exist. This is expected to change with PLATO  \citep{Rauer2014} which includes ground-based follow-up. By then, it will be possible to statistically assess trends on the mass-radius diagram \citep{Heller2022}.

\paragraph{Potential reasons from discrepancy of close-in ice rich planets.}
\label{sec:discrepancies}
While the synthetic and observed radii match well, we find a difference in the orbital period distribution of the close-in sub-Neptunes. In Fig. \ref{fig:pR_vanilla} by comparing the observed (left) with the synthetic, biased (middle) population, it becomes apparent that a group of sub-Neptunes is theoretically predicted to exist at orbital periods of \SIrange{1}{5}{d} which is absent in the observed sample. While they are only straddling the border of the observed paucity of Saturnian mass planets (the sub-Jovian desert \citep{Mazeh2016}) and the individual planets' radii are not in disagreement with observations, they are more numerous than observed. Instead, a similar number of planets is missing at longer orbital periods. It is thought \citep{Lopez2017} that even at high irradiation, steam envelopes can be kept for sufficient core masses. However, with the exception of 55 Cancri e \citep{Fischer2008}, there are no observed planets with low bulk densities in this regime. Thus, the very close-in sub-Neptune population that we obtain is disfavored by observations.

A possible explanation is that they might stop their migration in the type I regime \citep{Paardekooper2010,Paardekooper2011} at larger orbital distances than what the theoretical model currently predicts. At the moment, the inner edge of the gas disks acts as migration trap and is set to the co-rotation radius derived from observed rotation periods of young T Tauri stars \citep[e.g.][]{Henderson2011,Affer2013,Venuti2017}. However, the stopping distance could be further out. Since sub-Neptunes form early in the gaseous disks (to be able to migrate significantly), and given a sufficient source of disk turbulence in the inner disk, viscous heating would be efficient. Therefore, the location of ionization, that is, the inner edge of the dead zone to the magneto-rotational instability \citep{Gammie1996,Flock2016}, can lie further away from the star. This introduces a migration trap \citep{Mohanty2018,Ataiee2021} and causes the distribution of migrated planets to peak at larger orbits. In the future, the formation model needs to include the ionized region and a transition in the strength of magneto-rotational turbulence.
\begin{figure*}
	\centering
	\includegraphics[width=.825\linewidth]{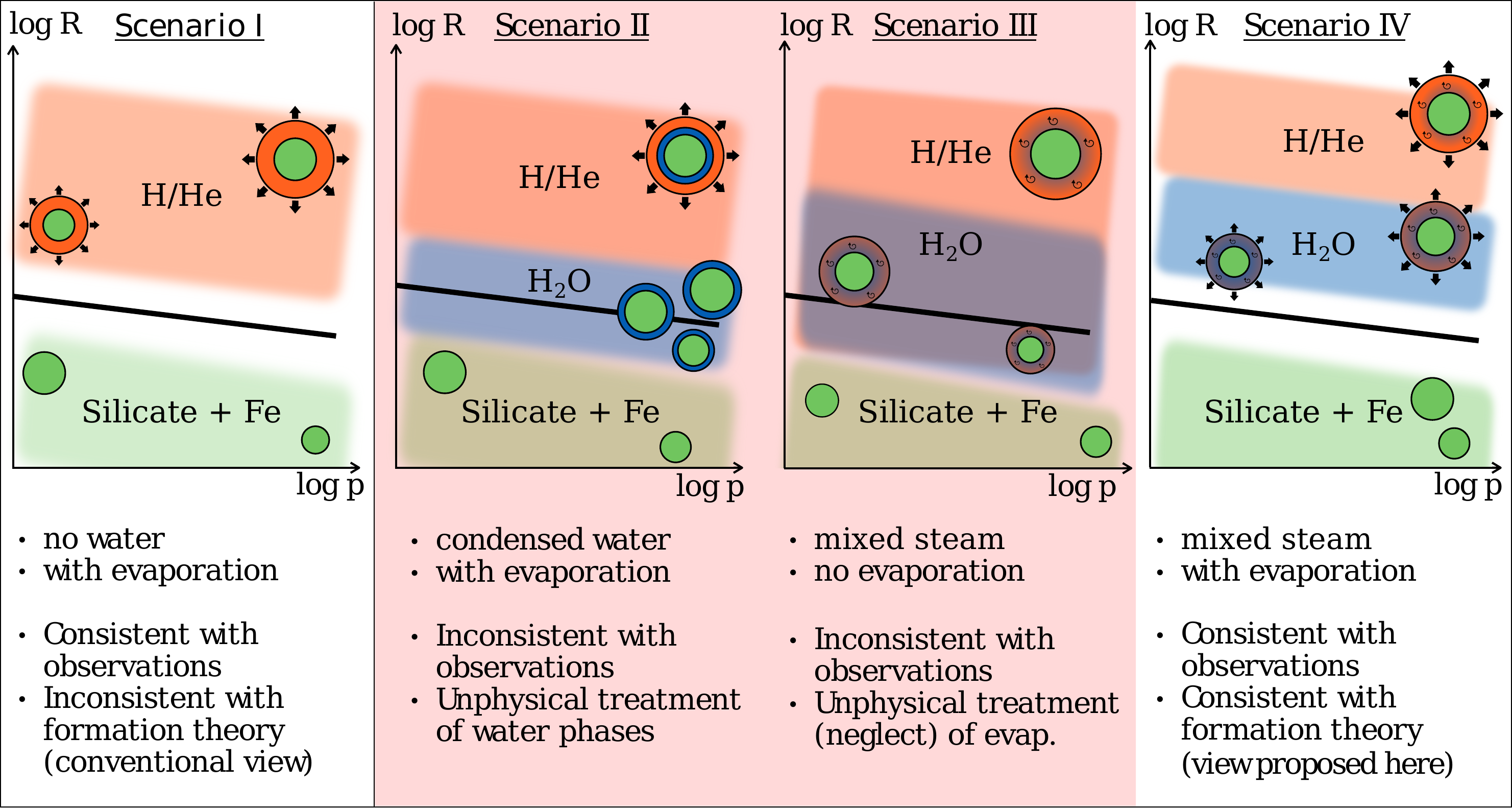}
	\caption{Schematic visualization of the four considered scenarios. The location of the observed radius valley is represented as black line while the planetary population of pure silicate and iron bodies (green), water-rich planets (blue), and H/He gas giants (red) are shown as blurred rectangles. Example interior structure schematics are shown with indicators for mixing and escape. We qualitatively found a negative slope of the distributions in all scenarios. We identified in this work that the radius valley is either shaped by a mass-loss mechanism without any water on the planets or by photoevaporation combined with water in supercritical steam envelopes which is in better agreement with the predictions of formation theory, in particular that planets should be undergoing orbital migration. Neither the scenario with condensed water (Scenario II) nor without photoevaporation (Scenario III) reproduce the observed valley.}
	\label{fig:schematic}
\end{figure*}

\section*{Summary and conclusion}
By coupling a global end-to-end planet formation model to different evolution pathways, we identify  scenarios which lead to a distribution of planetary radii consistent with the observed location of the radius valley \citep{Fulton2017}. These scenarios are shown in Figure \ref{fig:schematic}. Key mechanisms in the the formation model are orbital migration and \textit{N}-body interactions. The evolution part was calculated with and without updated models of photoevaporation and bloating. Furthermore, we assume in our nominal case that the accreted water is not condensed into solid ice on top of the rocky core as some previous works but is in the correct phase, which is typically supercritical vapor which can be mixed with H/He.

In this way, we provide theoretical support of the scenario of formation, or, more specifically, gas-driven orbital migration, shaping the distribution of mostly water-rich, steam-envelope planets populating the sub-Neptune peak (scenario IV in the schematic figure \ref{fig:schematic}). Only at larger planetary radii H/He becomes the dominant gaseous constituent. This is in agreement with an earlier study \citep{Venturini2020}, albeit the formation mechanism assumed in that study was pebble accretion. The bi-modal mass distribution (below \SI{20}{\mearth}) shaped by pebble accretion and its isolation mass is not required to match observed radii. Instead a uni-modal mass distribution can also reproduce the observations.

At the same time, we also give evidence for the necessity of an evolutionary mass-loss mechanism. Atmospheric escape is necessary to populate the super-Earth peak with evaporated rocky cores. Without evaporation, rocky planets with small H/He atmospheres which form during the gas disk stage inside of the ice-line would lead to low-mass planets with large radii resulting in a radius distribution inconsistent with observations (Scenario III in Fig. \ref{fig:schematic}).

Our results are not sensitive to modifying the photoevaporation model or the presence of a bloating mechanism. With both orbital migration and atmospheric escape causing the radius valley, one can speak of a hybrid origin of the radius valley caused by both formation and as evolution.

Assuming the mass distribution of our model is approximately correct, we can also falsify the scenario (II in Fig. \ref{fig:schematic}) of a condensed-out ice layer. In this scenario, the icy planets' radii fall into the valley and a lack of planets is found at the sub-Neptune peak. This echoes earlier works assuming condensed ice \citep{Owen2017,Jin2018}. However, at the planets' equilibrium temperatures, water is not in the solid ice form \cite{Turbet2020,Mousis2021,Pierrehumbert2023}.

Finally, based on planet evolution only we can not rule out the classical picture of photoevaporation of H/He envelopes on top of exclusively rocky cores \citep[][for a review]{Owen2019a} only shaping the radius distribution. However, as a key result, many different planet formation models consistently predict the migration of $\sim 10 M_\oplus$ water-rich planets to regions close to the host star, a prediction for example already made in the first generation of population syntheses\citep[][(their ``horizontal branch'' planets)]{Mordasini2009b}. This leads to the presence of a population of close-in, water-rich planets for which we showed that it will not lead to a disagreement with the observed valley. If the prediction of the presence of water-rich sub-Neptunes should turn out to be incorrect, it would call for a revision of fundamental aspects of formation theory. These aspects could for example be protoplanetary disk structures leading to less orbital migration \citep{Coleman2016a,Ogihara2018} or a high efficiency of volatile loss during planet assembly \citep{Lichtenberg2019}.

Today, direct observational constraints on the bulk composition of sub-Neptunes are still inconclusive but with the help of the James Webb Space Telescope and the future ARIEL mission \citep{Tinetti2018}, we expect to find more evidence for either the water-rich or the H/He dominated composition and advocate the investigation of sub-Neptune atmospheres to resolve the mysteries of the radius valley.

\appendix

\section*{Methods}
	\paragraph{Formation model}
	\label{app:formation_model}
	
	The formation part of the Bern model gathers the evolution of a viscous-accreting gas disk, the dynamical state of the solids in the disk, the concurrent accretion of solids and gas by the protoplanets, planet-disk interactions leading to gas-driven migration, and dynamical interactions between the protoplanets. Both the gas and solid disks are described by 1D axisymmetric, vertically-integrated profiles. The model solves an advection-diffusion equation to track the evolution of the gas disk \citep{Lust1952,Lynden-Bell1974}, with additional sink terms for the accretion by the protoplanets and external photoevaporation. The standard $\alpha$ parametrization \citep{Shakura1973} is used to compute the viscosity, while a radiative balance is used to determine the vertical structure \citep{Nakamoto1994}. Solids are assumed to be in planetesimals, whose dynamical state (eccentricity and inclination) is evolved due to damping by the gas disk, self-stirring, and stirring by the protoplanets \citep[following]{Fortier2013}.
	
	Planet formation follows the core accretion paradigm \citep{Perri1974,Mizuno1978,Mizuno1980} with the concurrent accretion of planetesimals and gas \citep{Pollack1996}. At time zero, 100 embryos of \SI{e-2}{\mearth} (nearly lunar-mass) are randomly placed in the disk with a uniform probability in the logarithm of the distance between the inner edge and \SI{40}{\au}. Planetesimal accretion is assumed to occur in the oligarchic regime \citep{Ida1993,Inaba2001,Thommes2003} and their radius is set to \SI{300}{\meter}. The enhancement of the capture cross-section by the presence of an envelope is included \citep{Inaba2003}. The model solves the 1D spherically symmetric internal structure equations \citep{Bodenheimer1986} to obtain either the mass or the radius of the envelope. During the early stages, gas accretion is limited by the cooling of the planet \citep{Pollack1996,LeeChiang2015} and the envelope mass is retrieved from the structure. Cooling efficiency improves as the planet grows, leading to a point where the gas accretion rate exceeds the supply from the disk, which is set according to the Bondi rate. Once this point is reached, the gas accretion is known and the internal structure equations are used to track the contraction of the envelope \citep{Bodenheimer2000}, yielding the planet radius and luminosity.
	
	The model prescribes gas-driven migration \citep{Paardekooper2011} plus a reduction of the co-rotation torques due to planet eccentricity and inclination \citep{Coleman2014}. Gas-induced eccentricity and inclination damping are also included \citep{Cresswell2008}. Dynamical planet-planet interactions are tracked using the \texttt{mercury} N-body code \citep{Chambers1999}.
	
	\paragraph{Improved evolution model}
	\label{app:evo_model}
	In the following, we describe in detail the changes in the evolution model compared to the previous version which was described in full detail \citep{Emsenhuber2020a}. The contraction and cooling of the interior structure as well as the tidal evolution remained unchanged.
	
	The first major change concerns the treatment of water in the interior structure model. Previously, we assumed that it is always in condensed form and resides in a pure water/ice shell between the inner iron-silicate core and a possible outer gas envelope which consists of pure H/He. In the New Generation Planetary Population Synthesis (NGPPS) series \citep{Emsenhuber2020a,Emsenhuber2020b}, this approximation was required as our interior structure model was not updated for mixed H/He + H$_2$O envelopes. Instead, we now allow for all phases of water -- including vapor and supercritical -- to exist. The orbital distances at which the valley is observed lies closer to the star than the habitable zone \citep{Kopparapu2013a}, meaning the temperatures at the top of the atmospheres exceed the threshold where water could condense out. Additionally, a strong runaway greenhouse occurs in vapor atmospheres at these distances \citep{Turbet2020,Pierrehumbert2023}. Given the fact that H$_2$O is also miscible in H/He on a molecular level \citep{Vazan2022,Pierrehumbert2023}. This means that in reality, water does not form a separate condensed layer, but is mixed into a H/He + H$_2$O vapor envelope (or in a pure vapor envelope if the planet contains no H/He). One of the latest updates to our interior structure and planet evolution model was to include such mixed compositions \citep{Linder2019}. Specifically, we mix water described with the AQUA equation of state (EoS, see also the description below) \citep{Haldemann2020} with H/He \citep{Chabrier2019}. \footnote{In NGPPS and therefore during the formation phase, an older EoS \citep{Saumon1995}  for H/He was used.} This implies that the fraction of heavy elements $Z$ is radially constant in the envelope. We note that, this averaged $Z$ is used to calculate molecular opacities as a function of temperature but assuming solar-like elemental abundances and equilibrium chemistry \citep{Freedman2014}.
	
	Second, and related to the previous step, we improved the prescription for X-ray and Extreme Ultra Violet (XEUV)-driven atmospheric escape of the planetary envelopes. XEUV radiation drives a mass loss rate of
	\begin{equation}
	\dot{M}_\mathrm{esc,EL} = \epsilon \frac{\pi F_\mathrm{XEUV} R_{\tau=2/3} R_\mathrm{base}^2}{G M_\mathrm{tot} K(\xi)}\,,
	\label{eq:escape}
	\end{equation}
	where $F_\mathrm{XEUV}$ is the flux received in either X-ray (dominating early) or EUV wavelengths, $R_\mathrm{base}$ is the base of the ionization layer, $R_{\tau=2/3}$ is the radius of the $\tau=2/3$ layer, $G$ the gravitational constant, $M_\mathrm{tot}$ the mass of the planet, $K(\xi) = 1 - \frac{3}{2 \xi} - \frac{1}{2 \xi^3}$, $\xi = R_\mathrm{Roche}/R_{\tau=2/3}$ the ratio of the planets Roche limit to its radius, and $\epsilon$ a escape efficiency factor. $R_\mathrm{base}$ is located in the planetary structure by equating the partial pressure to the pressure where an optical depth of one is reached for ultraviolet photons \citep{Murray-Clay2009}. If this critical pressure lies exterior to the resolved envelope structure, we extrapolate using the scale height determined at \SI{1}{bar}. As improvements to the NGPPS escape model \citep{Jin2018,Emsenhuber2020a}, the time evolution of $F_\mathrm{XEUV}$ is updated from a simple log-constant slope \citep{Ribas2005} to an tabulated model \citep{Mcdonald2019} based on more recent observational data \citep{Jackson2012,Shkolnik2014}. Furthermore, we calculate escape rates for both water and H/He separately. For water, we use Eq. \eqref{eq:escape} with a variable $\epsilon$ given such that the mass loss values found for a pure water vapor atmosphere \citep{Johnstone2020} are reproduced. In that work, the author makes use of a model \citep{Johnstone2018} which evolves hydrodynamically and chemically a 1D atmosphere from first principles. For the escape of H/He, we use extended tables \citep{Kubyshkina2021} based on a hydrodynamic model accounting for various heating (including XEUV) and cooling mechanisms \citep[described in][]{Kubyshkina2018}. The two rates from water and H/He escape are summed by weight $\dot{M}_\mathrm{esc,KuFo,Jo} =  \dot{M}_\mathrm{esc,H_{2}O,Jo} Z + \dot{M}_\mathrm{esc,H/He,KuFo} (1-Z)$, where $Z$ is the mass fraction of H$_2$O in the envelope. When removing mass from the envelope, we leave $Z$ unchanged, that is, we assume no fractionation which is motivated by classical theory for large escape rates \citep{Zahnle1990}.
	
	The third and final relevant variation is in the mechanism for so called 'bloating' where we use an observationally-derived, empirical relation \citep{Sarkis2021} in the nominal model. Bloating is the term used to describe an empirically found increase of planetary radii of mostly hot Jupiters over the expected theoretical values \citep{Guillot2002}. To reproduce observations, an additional heat source in the deeper interior of the planetary structure is required. We model this as an additional luminosity added to the energy equation at the core-envelope boundary. In addition to the empirical model \citep{Sarkis2021}, we also explore cases without bloating and with a fit (see the next paragraph) to the results of a physical model of potential temperature advection \citep{Tremblin2017}.

	\paragraph{Bloating model using a fit to the potential temperature advection model}
	\label{app:tremblin}
	
	To include potential temperature advection as a bloating mechanism \citep{Tremblin2017}, we start with a fit of the temperature at \SI{100}{bar} given a stellar irradiation flux $F$ \citep[their Fig. 5]{Tremblin2017}. We further subtract \SI{200}{\kelvin} motivated by the results from 3D simulations \citep{Sainsbury-Martinez2019a}. We first calculate the entropy, and then using our interior structure models the corresponding bloating luminosity $L_{\rm bloat}$.
		
	To take into account the mass fraction of heavy molecules (i.e. water) in the envelope $Z$, we use a $Z$ dependent fit of the form
	\begin{equation}
	L_\mathrm{bloat}(Z) = a(Z) + b(Z) \log_{10} F + c(Z) \log_{10}^2 F\,,
	\label{eq:L_of_F_fit}
	\end{equation}
	where $F$ is the irradiation flux. To obtain results for continuous values of $Z$ we interpolate linearly between the fits and we further found a linear dependency of the luminosity on planetary mass to obtain the desired entropy. The resulting parameters for equation \eqref{eq:L_of_F_fit} are listed in Table \ref{tab:L_of_F_fit}.
	
	\begin{table}[htbp]
		\caption{Fit parameters for Eq. \eqref{eq:L_of_F_fit} for a planet like HD209458b (\SI{241.9}{\mearth})}
		\label{tab:L_of_F_fit}
		\begin{tabular}{| r | c | c | c |}
			\hline
			Envelope enrichment  & $a$ & $b$ & $c$ \\ 
			\hline
			$Z = 0$ & -0.555 & -0.171 & 0.066\\
			$Z = 0.2$ & -9.581 &  1.881 &  -0.061 \\
			$Z = 0.4$ & -23.940 &  5.311 & -0.267 \\
			$Z = 0.6$ & -41.709 &  9.610 & -0.525 \\
			$Z = 0.8$ & -61.618 &  14.462 & -0.818 \\
			$Z = 1.0$ & -82.456 &  19.606 & -1.132 \\
			\hline
		\end{tabular}
	\end{table}

\begin{figure}
		\centering
		\includegraphics[width=\linewidth]{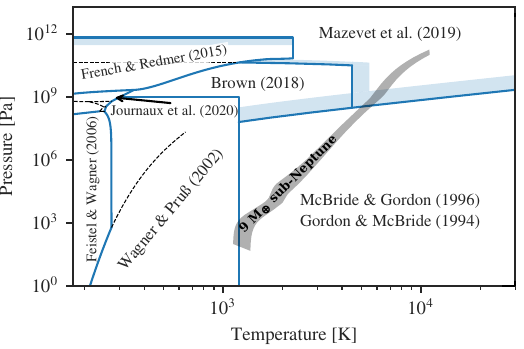}
		\caption{Phase diagram of H$_2$O split into the seven regions, separated by the solid blue lines. In each region a different EoS is used. Most region boundaries  follow phase transition curves of H$_2$O. The dashed lines are phase transitions that are not region boundaries, meaning the same EoS is used along the phase transition. The blue shaded areas show where neighboring regions have to be interpolated to achieve smooth transitions in the state parameters. The time evolution of a sub-Neptune water envelope at \SI{3}{days} orbital period with an initial mass of \SI{2.7}{\mearth} on top of a rocky \SI{6.3}{\mearth} core is shown in gray. Its inner layers are supercritical at high pressure \citep{Mazevet2019} and the upper layers mostly in the high temperature gas regime \citep{cea1_1994,cea2_1996}.}
		\label{fig:aqua}
	\end{figure}

	\paragraph{AQUA equation of state}
	\label{sec:aqua}
	The AQUA equation of state is a collection of seven individual equations of state of H$_2$O, which together cover a large domain in pressure and temperature useful to model planetary interiors \citep{Haldemann2020}. H$_2$O is a very peculiar molecule which shows a large number of distinct solid phases and a complex phase diagram. Given its significance in industrial applications, H$_2$O is well described by EoS at low pressures, i.e. $P<1$ GPa \citep{wagner_iapws_2002,feistel_new_2006,wagner_2011}.
	A single EoS which incorporates the many phases of H$_2$O and covers the necessary large range in pressure and temperature has not yet been published. All commonly used H$_2$O-EoS which cover a large range in pressure and temperature, make significant simplifications in terms of the number of treated phases and the location of the phase transitions. AQUA thus combines seven state of the art EoS into a single tabulated EoS which covers a domain from 0.1 Pa to 400 TPa in pressure and 150 K to $10^5$ K in temperature.
	Each EoS is used in a distinct region in P-T space and the included phases are (a) ice-Ih\citep{feistel_new_2006}; (b) ice-II, ice-III, ice-V, and ice-VI\citep{journaux_holistic_2020}; (c) ice-VII, ice-VII*, and ice-X\citep{french_redmer_2015}; (d) low temperature gas, low-pressure liquid and low-pressure supercritical fluid\citep{wagner_iapws_2002}; (e) higher-pressure supercritical fluid  \citep{brown_local_2018}; (f) high-temperature gas including ionization and dissociation of H$_2$O \citep{cea1_1994,cea2_1996}; and (g) supercritical fluid at very high pressures including super-ionic phases \citep{Mazevet2019}.
	The location of the seven regions from individual EoS are shown in Fig. \ref{fig:aqua}. Since there are region boundaries which do not follow a physical phase transition, e.g. between \citep{brown_local_2018} and \citep{Mazevet2019}, AQUA provides interpolated values, to assure a smooth transition in all provided state parameters. The state parameters which AQUA provides for a given pressure and temperature, are density $\rho$, adiabatic gradient $\Delta_\text{Ad}=\left(\frac{\partial \ln T}{\partial \ln P}\right)_S$, entropy $s$, internal energy $u$, bulk speed of sound $w$, mean molecular weight $\mu$, ionization fraction $x_\text{ion}$, dissociation fraction $x_d$, and a phase identifier to identify the corresponding phase.
	
	\paragraph{KOBE}
	\label{sec:KOBE}
	\texttt{Kepler Observes Bern Exoplanets} (KOBE) \citep{Mishra2021} is a program that simulates transit surveys of exoplanets \footnote{KOBE is publicly available at \url{https://github.com/exomishra/kobe}}. If Kepler (or TESS, PLATO, etc.) was to, hypothetically, observe a synthetic population of planets then KOBE identifies those synthetic planets that would have been detected by such a  survey.
	As a result, KOBE allows theoretical models of planet formation (such as the Bern Model used here) to be compared with transit observations. 
	
	KOBE is organized in three sequential modules. The first module, KOBE-Shadow, finds transiting planets from a synthetic population of planets given their orbital elements. A planet can only transit when its orbit is aligned with respect to a hypothetical observer's line-of-sight. KOBE-Transits, the next module, calculates transit parameters. It applies the detection biases coming from physical limitations; large planets in tight orbits around a quiet star are strongly favored. For comparison to Kepler, transiting planets which transit at least three times and have a signal to noise ratio $\ge 7.1$ are potentially detectable. KOBE-Vetter, the final module, applies the completeness and reliability of the Kepler pipeline by emulating Kepler's Robovetter \citep{Thompson2018}. Transiting planets that are identified as planetary candidates by KOBE-Vetter make up the KOBE-biased catalog which we use for comparison to the Kepler-based CKS results \citep{Petigura2018}. The planets in KOBE catalog are comparable to the exoplanet population discovered by Kepler.

\section*{Acknowledgments}
R.B. and Th.H. acknowledge support from the European Research Council under the European Union’s Horizon 2020 research and innovation program under grant agreement No. 832428-Origins. C.M. and A.E. acknowledge the support from the Swiss National Science Foundation under grant 200021\_204847 ``PlanetsInTime''. Parts of this work has been carried out within the framework of the NCCR PlanetS supported by the Swiss National Science Foundation under grants 51NF40\_182901 and 51NF40\_205606. Calculations were performed on the Horus cluster at the University of Bern.

\section*{Author contributions}
R.B. analyzed the synthetic data and wrote the manuscript with contributions by all other authors. C.M., L.M, R.B., and A.E. ran the numerical experiments. C.M. and A.E. are lead developers of the numerical code. L.M. developed the code for comparing theory with observations. The updated Equation of State was integrated by J.H., J.V., and C.M.. C.M. conceived the original idea. Th.H. and C.M. supervised the project.

\section*{Competing interests}
The authors declare no competing interests.

\section*{Data availability}
Raw data used in this study are accessible at \href{https://dace.unige.ch/}{dace.unige.ch} under identifier NG76. Derived data supporting the findings of this study, including source data for Figures 1,2, and 3, and Extended Data Figures 7, 8, and 9, are publicly available after publication of the manuscript. Supplementary data is available upon reasonable request.

\section{Extended Data}
\vspace{400px}
\begin{figure}[htbp]
	\centering
	\includegraphics[width=\linewidth,trim= 4 33 0 7 ,clip]{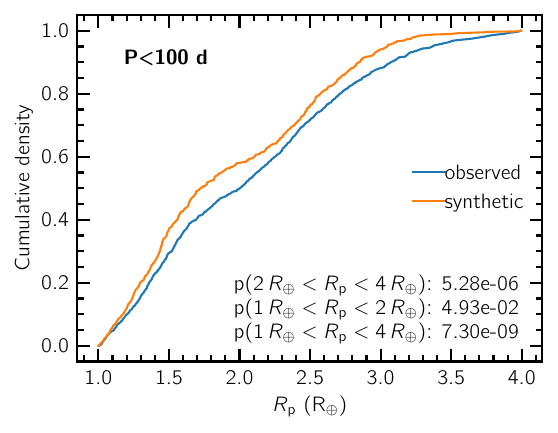}
	\includegraphics[width=\linewidth,trim= 4 33 0 7 ,clip]{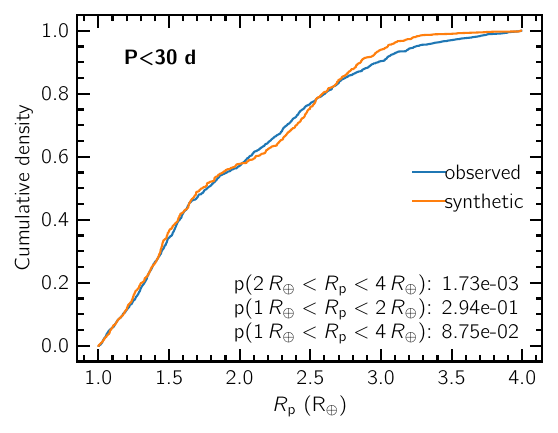}
	\includegraphics[width=\linewidth,trim= 4 5 0 7 ,clip]{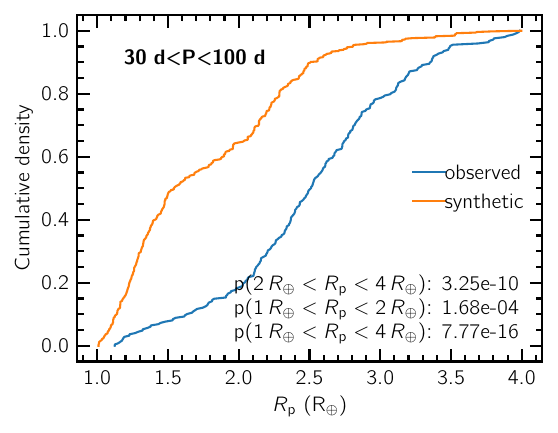}
	\caption{Cumulative radius distributions of observed and synthetic super-Earth and sub-Neptune planets from the nominal simulation in different orbital regions. Two-sample, Kolmogorov-Smirnov (KS) test p-values are listed in the bottom right. The null hypothesis that the observed and the synthetic distributions are the same can be rejected for the region outside of \SI{30}{day} orbital periods and consistently for sub-Neptunes. However, for super-Earths the two distributions do not disagree significantly (p-values above 0.05) which even leads to a p-value above 0.05 for the whole radius range if only planets on short orbits are considered. }
	\label{fig:statistics}
\end{figure}

\begin{figure}[htbp]
	\centering
	\includegraphics[width=\linewidth,trim= 4 5 0 7 ,clip]{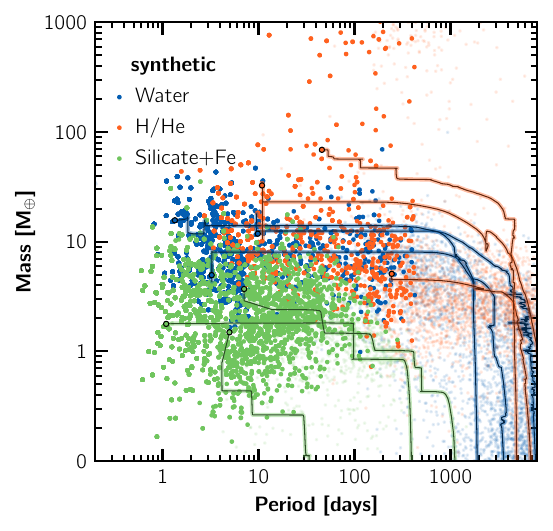}
	\caption{Orbital period against masses of synthetic planets after \SI{5}{Gyr} of evolution. Planets observable and non-observable with the Kepler space telescope are shown (KOBE bias applied) as opaque or transparent points respectively. The points are colored as in Fig. \ref{fig:pR_vanilla} and we show in addition the formation tracks of a selection of planets from the different categories as lines. Stationary jumps in mass are caused by giant impacts. Since there is no effect of the lower density material on the planetary mass, the radius valley is not visible in mass space. 
	}
	\label{fig:pM_vanilla}
\end{figure}

\begin{figure}
	\centering
	\includegraphics[width=\linewidth]{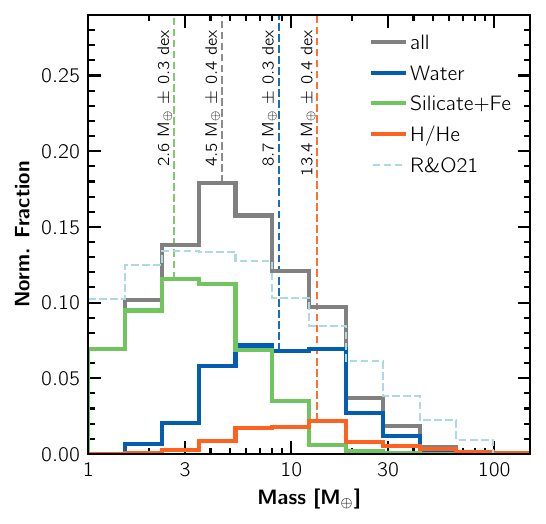}
	\caption{Mass histogram of the nominal simulation with applied Kepler bias. Lines show different groups: all planets (grey), rocky planets (only iron core and silicate mantle, green), planets with water but no H/He (blue), and planets with H/He (red). Dashed lines mark logarithmic mean values of each category and are labeled with mean and standard deviations. The distribution of core masses retrieved by Rogers \& Owen \citep{Rogers2021a} (R\&O21) in a pure photoevaporation scenario is shown for comparison.}
	\label{fig:mass_hist}
\end{figure}

\begin{figure}
	\centering
	\includegraphics[width=\linewidth]{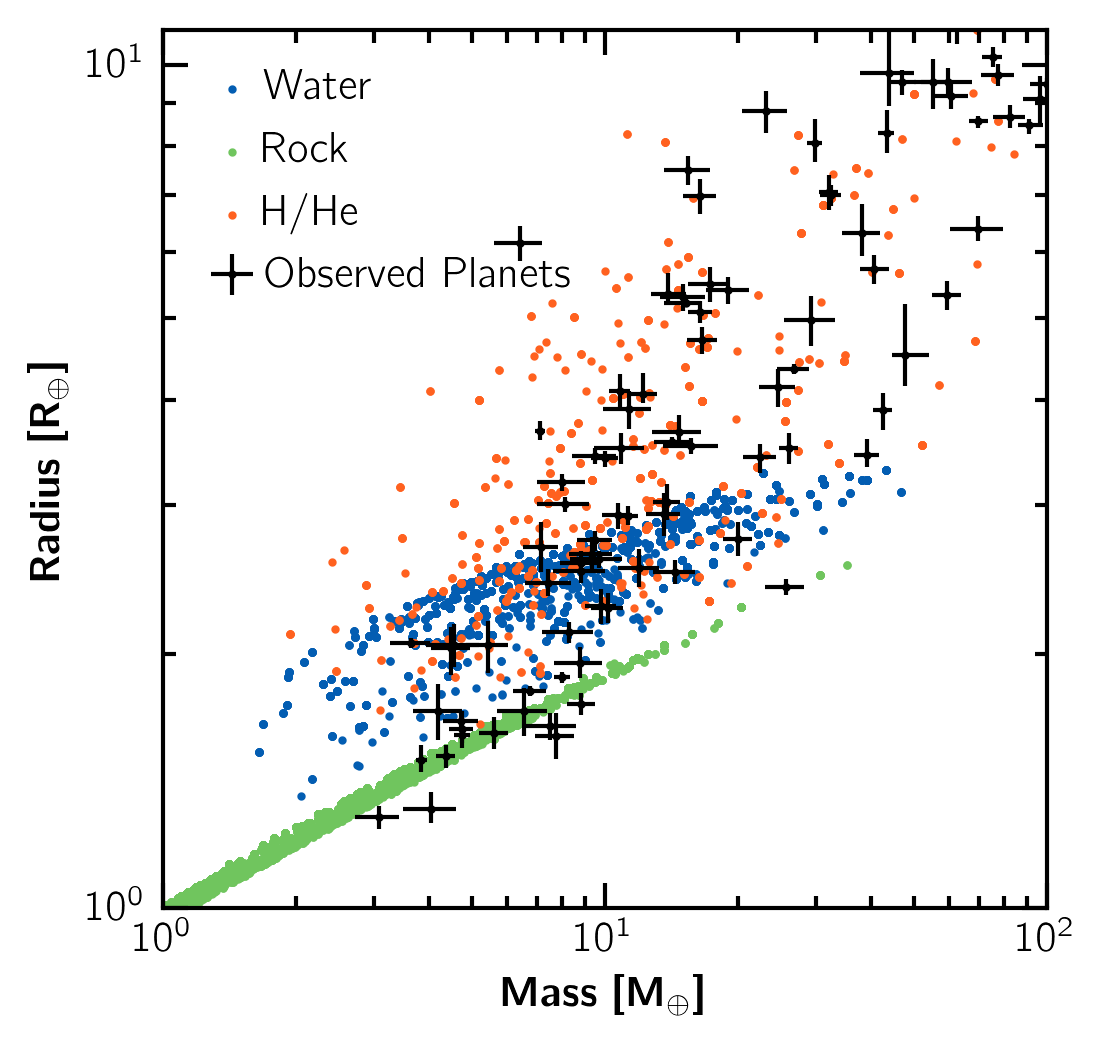}
	\caption{Total planetary masses against planetary radii of observed and synthetic planets. The observational data was taken from the NASA Exoplanet Archive (composite data, accessed on Dec 9, 2022) filtering for relative mass, stellar mass, and radius standard errors smaller than \SI{20}{\percent} and stellar masses ranging from \SIrange{0.75}{1.25}{M_{\odot}}. Errorbars show the standard error. We caution that this sample of planets does not come from a homogeneous survey.
	}
	\label{fig:mR}
\end{figure}

\pagebreak
	\section{Supplementary Information}
\subsection{Dependency on model assumptions}
We investigate the effect of several key model assumptions on the final planet properties. We use the same planetary population after the formation stage but change the used description for the effects of photoevaporation and bloating of the planetary envelopes.

\begin{figure*}
	\centering
	\includegraphics[width=\linewidth,trim=5 5 5 5, clip]{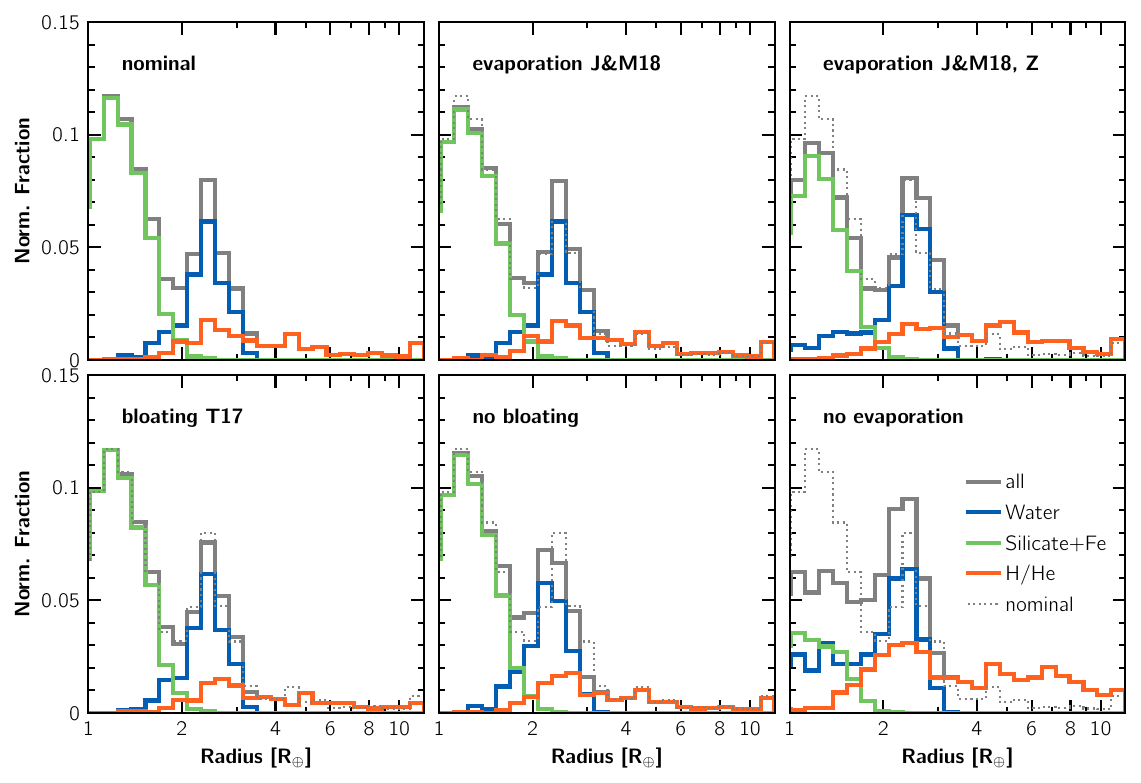}
	\caption{Histogram of planetary radii for different assumptions in the evolution model. All planets with orbital periods less than \SI{100}{d} are included. The colored histogram are the same as in Fig. 2.}
	\label{fig:evap_bloat}
\end{figure*}

\begin{figure*}
	\centering
	\includegraphics[width=\linewidth,trim=5 5 5 5, clip]{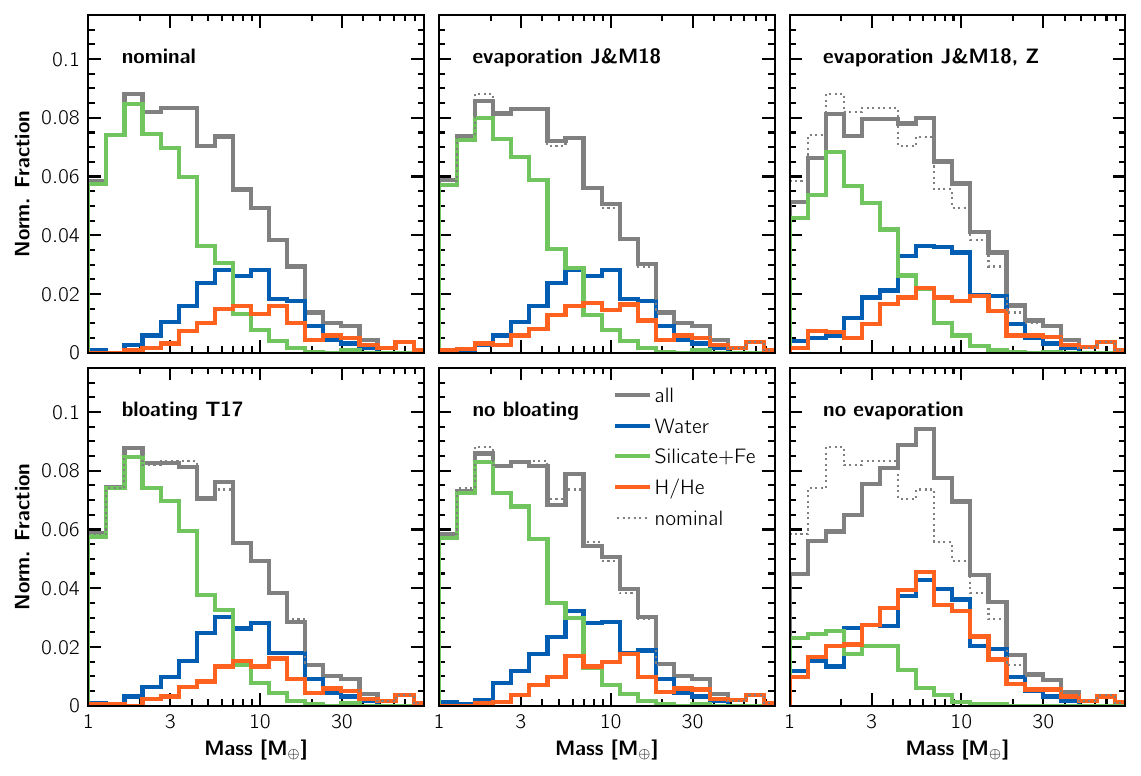}
	\caption{Histogram of planetary masses for different assumptions in the evolution model. The selection of planets and colors are identical to Supp. Fig. \ref{fig:evap_bloat}. }
	\label{fig:evap_bloat_masses}
\end{figure*}

\subsubsection{Impact of the evaporation model}
As a first test, we see that removing the effect of photoevaporation results in a distribution of radii without a valley. Photoevaporation is a necessary process to populate the envelope-stripped, super-Earth peak (see Fig. 2 in the main text) with planets located close to or within the observed radius valley otherwise. Furthermore, without photoevaporation, planets in the super-Earth regime would commonly contain large amounts of water, which is not realistic at these planetary masses and distances to the star \citep{Lopez2017}.

To model photoevaporation, we nominally used the tabulated evaporation rates based on recent, detailed radiation-hydrodynamic models for H/He \citep{Kubyshkina2021} and water \citep{Johnstone2020} dominated envelopes. To contrast them to a more simple model, we show in Supp. Fig. \ref{fig:evap_bloat} in the upper, middle panel the resulting distribution of planetary radii using energy- and radiation-recombination-limited escape rates \citep{Jin2014,Jin2018} and further a variant thereof which accounts for an increase in mean molecular weight as a function of the envelope metallicity $Z$ when calculating the base layer of the evaporative flow and a scaling of the efficiency factor $\epsilon$ with $Z$ in Eq. 1 in Methods taken from ref. \citep{Lopez2017}. While the former gives results similar to the nominal evaporation model, the $Z$-dependent version results in a less pronounced super-Earth peak as less planets lost their primordial envelope.

For reference, we also show the resulting planetary masses in Supp. Fig. \ref{fig:evap_bloat_masses}. A shifted overall peak at higher masses is found in mass space when removing the effect of photoevaporation but no significant differences result based on the different evaporation model. 

\subsubsection{Impact of bloating}
\label{app:bloating}
In Supp. Fig. \ref{fig:evap_bloat}, we show panels with the nominal, empirical bloating model \citep{Sarkis2021} (top left), with radius inflation caused by the potential temperature advection mechanism\citep{Tremblin2017} (bottom, left), and without any bloating mechanism at all (bottom, middle). If included, we assume that the mechanism is acting on all planetary masses and we see a small impact of the bloating mechanism on the sub-Neptune population (blue histogram). It is expected that bloating by ohmic dissipation is efficient in sub-Neptunes \citep{Pu2017}. Without a bloating mechanism the valley would be more populated by smaller water-rich planets therefore making it less deep and visible. The effect is however too small to be constrained by the current observational data. Both compared bloating models give very similar results. Overall, we find that the presence of a radius valley is robust against uncertain extrapolations of the radius inflation mechanism to lower planetary masses and different compositions.

\bibliographystyle{naturemag}
\bibliography{library_betterbib_zotero,add} 

\end{document}